\documentclass{svjour3}                     
\smartqed  
\usepackage{graphicx}
\journalname{Few-Body Systems (FB20)}
\begin{document}

\title{
A new EFT approach to NN scattering problem
\thanks{Presented at the 20th International IUPAP Conference on Few-Body Problems in Physics, 20 - 25 August, 2012, Fukuoka, Japan}
}


\author{E.~Epelbaum         \and
        J.~Gegelia 
}


\institute{E.~Epelbaum \at
             Institut f\" ur Theoretische Physik II, Fakult\" at f\" ur Physik und Astronomie,\\ Ruhr-Universit\" at Bochum 44780 Bochum, Germany \\
              Tel.:  	+49 (0)234 32-23707\\
              Fax:  	+49 (0)234 32-14697\\
              \email{evgeny.epelbaum@ruhr-uni-bochum.de}           
           \and
           J.~Gegelia \at
              Institut f\" ur Theoretische Physik II, Fakult\" at f\" ur Physik und Astronomie,\\ Ruhr-Universit\" at Bochum 44780 Bochum, Germany,\\
              Tbilisi State University, 0186 Tbilisi, Georgia\\
              Tel.:  	+49 (0)234 32-28957\\
              Fax:  	+49 (0)234 32-14697\\
              \email{Jambul.Gegelia@tp2.ruhr-uni-bochum.de}
}

\date{Received: date / Accepted: date}

\maketitle

\begin{abstract}
We present the new, modified Weinberg approach to the NN scattering problem in effective field theory.
Issues of renormalization are briefly discussed and the results of LO calculations are presented.
\end{abstract}

\section{Introduction}
\label{intro}

The few-nucleon sector of baryon chiral perturbation theory was first considered in Ref.~\cite{Weinberg:rz} (For a recent review see e.g. \cite{Epelbaum:2008ga}). Within Weinberg's approach, the  nucleon-nucleon (NN) potential is defined as a sum of all two-nucleon-irreducible time ordered diagrams of the non-relativistic effective field theory (EFT). It is calculated as a systematic expansion in small parameters.
A finite number of diagrams contribute in the effective potential up to any finite order. The scattering amplitude is obtained by solving the Lippmann-Schwinger (LS) equation
\begin{equation}
{T(\vec p\,',\vec p,k)=V(\vec p\,',\vec p)+\hbar\, m \int \frac{d^3\vec q}{(2\,\pi)^3} \, V(\vec p\,',\vec q)\,
\frac{1}{\vec p^2-\vec q^2+i\,\epsilon}\, T(\vec q,\vec p,k),}
\label{IntEq}
\end{equation}
where ${\hbar}$ is included to keep the trace of the loop integration.
The leading order NN potential is given by
\begin{equation}
V_0 = C_S+C_T\, \vec\sigma_1\cdot\vec\sigma_2 -\frac{g_A^2}{4\,F^2}\ \vec\tau_1\cdot\vec\tau_2 \ \frac{\vec\sigma_1\cdot\vec q\,\vec\sigma_2\cdot\vec q}{q^2+M_\pi^2},
\label{LOV}
\end{equation}
with standard notations.

\section{Renormalization}
\label{sec:1}
In EFT, all divergences appearing in loop diagrams are absorbed in parameters of the effective Lagrangian. While non-perturbative
expressions may contain pieces which do not contribute in perturbative series, perturbative expansion of properly renormalized EFT
amplitudes in the region of applicability of perturbation theory must reproduce the properly renormalized perturbative amplitudes.

\subsection{How {\bf not} to renormalize in EFT}

To address the issue of non-perturbative renormalization in nuclear EFT
we consider a simple case of the LS equation with a contact interaction potential
$$
{ V(p',p)=C+C_2 (p\,'^2+p^2).}
$$
The corresponding LS equation can be solved exactly leading to \cite{daniel}
\begin{eqnarray}
{T} & {= }& {\frac{C+C_2^2
\hbar I_5+ k^2C_2\left( 2-C_2 \hbar I_3\right)}{\left( 1-C_2 \hbar
I_3\right)^2-\left[C+C_2^2 \hbar I_5+ k^2 C_2\left( 2-C_2 \hbar
I_3\right)\right] \hbar I(k)}\,,} \label{2x}
\end{eqnarray}
where ${I(k)}$, ${I_3}$ and ${I_5}$ are divergent one-loop integrals.
The parameters ${C}$ and ${C_2}$ can be determined by matching to the effective range expansion
$$
{\Re \left( T^{-1} \right) = - \frac{m}{4 \pi} \left(
-\frac{1}{a} +
  \frac{1}{2} r k^2 + \ldots \right) \,,}
$$ yielding
\begin{eqnarray}
{C }& {= }& {\frac{6 \pi ^2 [
a^2 \hbar \Lambda^3 m (64 \hbar -3 \pi \Lambda r) -6 \left(D -3 \pi
^2 \Lambda
  m\right) -62 \pi  a \hbar\Lambda^2 m ]
}{5 \hbar \Lambda^2 m^2 \left[a^2 \hbar \Lambda^2 (16  \hbar -\pi
\Lambda
 r)-12 \pi a \hbar \Lambda + 3 \pi ^2 \right]}\,,} \nonumber\\
{C_2 }& {= }& { -\frac{6 \pi
^2 [ - D + a^2
 \hbar m \Lambda ^3 (16 \hbar - \pi r \Lambda )-12 \pi
   a \hbar m \Lambda ^2 + 3 \pi ^2 m \Lambda ]
}{\hbar  m^2 \Lambda ^4 \left[a^2 \hbar \Lambda ^2 (16
   \hbar - \pi r \Lambda )-12 \pi  a \hbar  \Lambda +3 \pi
   ^2\right]} \,,}
\nonumber
\end{eqnarray}
where
$$ { D = \sqrt{3} \sqrt{\Lambda^2  m^2 (\pi -2
   a \hbar\Lambda)^2 \left(a^2 \hbar \Lambda^2 (16 \hbar -\pi
     \,\Lambda
  r)-12 \pi a \hbar \Lambda + 3 \pi ^2\right)}}\,.
$$
Using these expressions the scattering
amplitude can be expressed in terms of ${a}$ and ${r}$
\begin{eqnarray}
{T^{-1}} &{=}& {\frac{m }{4
\pi ^2 a \left[a \left(\pi  k^2 \,r\, \Lambda -4 \,\hbar\,
   \left(k^2+\Lambda ^2\right)\right)+2 \pi  \Lambda \right]}} \Biggl\{2
\Lambda  \biggl[a^2 \,\hbar\, k^2 (\pi \,r\, \Lambda -4 \,\hbar)-2
\pi  a \,\hbar\,
\Lambda \nonumber\\
&& +\pi ^2\biggr]{+a \,\hbar\, k \ln \frac{\Lambda -k}{\Lambda +k}
\left[a \left(\pi
   k^2 \,r\, \Lambda -4 \,\hbar\, \left(k^2+\Lambda
   ^2\right)\right)+2 \pi  \Lambda \right]\Biggr\}} +  i \hbar \frac{m k}{4 \pi }\,.
\label{invamplCutOff}
\end{eqnarray}
The amplitude ${T^{-1}}$ is finite for ${\Lambda \to
\infty}$ resulting in
$${T^{-1} = - \frac{m}{4 \pi} \left( -
\frac{1}{a} + \frac{1}{2} r k^2 - i \hbar\, k \right) + \mathcal{O}
\left(\Lambda^{-1} \right).}
$$
However, in the loop expansion of the ''renormalized'' amplitude of Eq.~(\ref{invamplCutOff})
\begin{eqnarray}
{T }&{ =}&  {\frac{2 \pi  \,a^2 k^2 \,r}{m}+\frac{4 \pi
   \,a}{m} -\frac{i \,\pi  \,a^2 \hbar\, k \left(a k^2\,r+2\right)^2}{m} }\nonumber\\
&{+}& {\hbar \frac{2 m\,k^4}{\pi^2} \Bigg[
-\frac{2 \,a^4\, k^4 {\Lambda}
   \,r^2}{m} + \frac{2 \left(a^4\, k^6 \,r^2+4 \,a^3\, k^4 \,r\right)}{\Lambda  \,m} + \mathcal{O}
\left(\Lambda^{-2} \right) \Bigg]} + {\mathcal{O}
\left(\hbar^{2}\right)} \nonumber
\end{eqnarray}
we see that not all divergences are removed. 
Therefore, the above ''non-perturbative renormalization'', although it gives a finite result, is not an EFT renormalization.
This procedure is more in spirit of ''peratization'' \cite{Feinberg,Epelbaum1}.
Not surprisingly, EFT using the ''non-perturbative renormalization'' fails to describe the phase shifts even at ${\rm N^3LO}$ \cite{Zeoli}.

\section{Modified Weinberg's approach}
\label{sec:2}

Equation (\ref{IntEq}) with the LO potential of Eq.~(\ref{LOV}) is not renormalizable, i.e. not all the divergences appearing
in iterations of the LS equation can be absorbed into redefinition of parameters of $V_0$.
This problem is often referred to as ''inconsistency of Weinberg's approach'' \cite{Savage:1998vh}.

A modified, renormalizable version of Weinberg's approach has been proposed in Ref.~\cite{Epelbaum:2012ua}.
We start with the manifestly Lorentz invariant effective Lagrangian of interacting pions and nucleons
and use time ordered perturbation theory \cite{Weinberg1}.
The effective NN potential $V$ is defined as a sum of two-nucleon-irreducible diagrams.
The off-shell nucleon-nucleon scattering amplitude ${T}$ satisfies the integral equation
$$
{T = V + V \,G \ T \,,}
$$
where ${G}$ is the two-nucleon propagator.
We expand $V$, $G$ and $T$ in small parameters,
\begin{eqnarray}
{ T}  {=}  {T_0 +T_1 +T_2 +\cdots\,,} \ \ \
{ G}  {=}  {G_0 +G_1 +G_2 +\cdots\,,} \ \ \
{ V}  {=}  {V_0 +V_1 +V_2 +\cdots\,,}
\end{eqnarray}
and  solve ${T}$ order by order. The LO amplitude is obtained by solving the equation
$$
{T_0 = V_0 + V_0\, G_0\, T_0\,.}
$$
Using ${T_0}$, the next-to-leading order (NLO) amplitude is obtained as
%
$$
{T_1=V_1+ T_0\,G_0\, V_1+ V_1\,G_0\, T_0+ T_0\,G_0\,V_1\,G_0\,T_0+
T_0\,G_1\, T_0\,.}
$$
Further, using ${T_0}$ and ${T_1}$, we calculate the NNLO amplitude ${T_2}$
etc.

The leading-order equation in the center-of-mass frame has the form
\begin{equation}
{T _{0 }\left(
\vec p\,',\vec p\right)}{=}{ V _{0 }\left(
\vec p\,',\vec p\right) - \frac{m^2}{2}\, \int \frac{d^3 \vec k}{(2\,\pi)^3}} {\frac{V _{0 }\left(
\vec p\,',\vec k\right) \, T _{0 }\left(
\vec k,\vec p\right)}{\left(\vec k^2+m^2\right)\, \left(
p_0-\sqrt{\vec k^2+m^2}+i\,\epsilon\right)},}
\label{kad}
\end{equation}
where $p_0=\sqrt{\vec p^2+m^2}$ is the energy of a single nucleon.
This equation was first obtained in Ref.~\cite{kadyshevsky}.
Its iterations generate only overall logarithmic divergences.
Therefore, the LO equation is perturbatively renormalizable.

To investigate the  non-perturbative regime, we notice that the ultraviolet behavior of Eq.~(\ref{kad})
corresponds to the potential ${\sim 1/r^2}$ for ${r\to 0}$ in ${2+1}$ dimensions.
More singular ${\sim 1/r^3}$ ultraviolet behavior
in non-relativistic EFT  is an artifact of that formulation.
    { This can be easily understood for the cutoff-regularized one-loop
integral 
\begin{equation}
{I}  = { \frac{4\,i}{(2\,\pi)^4}\int \frac{d^4 k \,\theta(\Lambda - |\vec k|)}
{\left[k^2-m^2+i\,0^+\right]\left[(P-k)^2-m^2+i\,0^+\right]}},
\end{equation}
%
%
%
where ${P=(2 \sqrt{m^2+ \vec p\,^2},\vec 0\,)}$.
The result of the above integral for ${\Lambda> |\vec p|}$ reads:
\begin{eqnarray}
{I} &{=}& {\frac{|\vec p|}{\pi ^2
   \sqrt{m^2+\vec p\,^2}} \ln \frac{\Lambda  \sqrt{m^2+\vec p\,^2}+|\vec p| \sqrt{\Lambda
   ^2+m^2}}{m \sqrt{\Lambda ^2-\vec p\,^2}}-\frac{\ln \frac{\Lambda +\sqrt{\Lambda
   ^2+m^2}}{m}}{\pi ^2}-\frac{i |\vec p|}{2 \pi\sqrt{m^2+\vec p\,^2}}\,.}
   \nonumber
\end{eqnarray}
Expanding first in ${1/\Lambda}$ and subsequently in ${1/m}$ we obtain
\begin{eqnarray}
{I} &{ =} & {-\frac{i |\vec p|}{2 \pi  m}  -\frac{{\ln\frac{\Lambda}{m}}}{\pi ^2}-\frac{\ln 2}{\pi ^2}+{\cal O}\left(\frac{1}{\Lambda^2},\frac{1}{m^2} \right).}
\label{fexp}
\end{eqnarray}
On the other hand, expanding first in ${1/m}$ and then in ${1/\Lambda}$ leads to:
\begin{eqnarray}
{I} & {=} & {-\frac{i |\vec p|}{2 \pi  m}-\frac{{\Lambda} }{\pi ^2 m}+
\frac{\vec p\,^2}{\pi ^2 \Lambda  m}+{\cal O}\left( \frac{1}{m^2},\frac{1}{\Lambda^2}\right).}
\label{sexp}
\end{eqnarray}
The non-relativistic approach corresponds to the second expansion.
It leads to the qualitatively different UV behavior.
In perturbative calculations one compensates for this mismatch by readjusting terms from the effective non-relativistic Lagrangian \cite{Gegelia:1999gf}.
When re-summing  iterations to all orders (e.g. by solving integral equations), one needs to include contributions of an infinite number of counter terms.
Otherwise, one is not allowed to take   ${\Lambda>m}$ in non-relativistic approach.
On the other hand, the integral
$${I_1=  {\frac{1}{(2\,\pi)^3}\int \frac{d^3 \vec k \,\theta(\Lambda - |\vec k|)}
{\left[\vec k^2+m^2\right]\left[p_0-\sqrt{\vec k^2+m^2}+i\,0^+\right]}}}$$
corresponding to Eq.~(\ref{kad}), has the proper UV behavior, guaranteeing the correct qualitative UV behavior in the modified Weinberg approach.}

The leading-order partial wave equations are found to have unique solutions, except for the ${^3P_0}$ wave.
The equation in the ${^3P_0}$ partial wave, similarly to the Skornyakov-Ter-Martirosyan equation \cite{skorniakov}, does not have an unique
solution. Analogously to Ref.~\cite{bedaque}
we solve this problem by including a counter-term   ${\frac{c(\Lambda)\, p\, p'}{\Lambda^2}}$ at leading order.

The phase shifts obtained from the LO calculations are in a good agreement (within the accuracy expected at LO) with
the Nijmegen PWA  \cite{deswart} (see figures).

The proper inclusion of the pion-exchange physics
can also be tested in predictions for the coefficients of the effective range expansion
\begin{equation}
p^{2l+1} \cot \delta_l (p) = - \frac{1}{a} + \frac{1}{2} r p^2 + v_2 p^4
+ v_3 p^6 + v_4 p^8 + \ldots\,,
\end{equation}
where $a$, $r$ and $v_i$ denote the scattering length, effective range
and shape parameters for the orbital angular
momentum $l$.
The long-range tail of the interaction imposes correlations between the
coefficients in the effective range expansion \cite{Cohen:1998jr}.
These correlations may
be regarded as low-energy theorems (LETs). In tables \ref{tab:1} and \ref{tab:2}, the LETs in the  KSW and modified Weinberg
approaches are compared with the results of the Nijmegen PWA
for the $^1S_0$ and $^3S_1$ partial waves, respectively.
Coefficients corresponding to Nijmegen PWA are taken from Refs.~\cite{pavon}, \cite{deswart}.
\begin{table}
\caption{Coefficients in ERE of the ${^1S_0}$ phase shifts}
\label{tab:1}       
\begin{tabular}{llllll}
\hline\noalign{\smallskip}
$^1S_0$ partial wave& $a$  [fm] & $r$  [fm] & $v_2$  [fm$^3$]& $v_3$  [fm$^5$]  & $v_4$  [fm$^7$]  \\
\noalign{\smallskip}\hline\noalign{\smallskip}
NLO KSW & fit   & fit  & $-3.3$  & $18$ & $-108$\\
LO Weinberg  &  fit  & $1.50$  & $-1.9$ &$8.6(8)$  &$-37(10)$\\
Nijmegen PWA & $-23.7$  & $2.67$  & $-0.5$  & $4.0$ & $-20$\\
\noalign{\smallskip}\hline
\end{tabular}
\end{table}
\begin{table}
\caption{Coefficients in ERE of the ${^3S_1}$ phase shifts}
\label{tab:2}       
\begin{tabular}{llllll}
\hline\noalign{\smallskip}
$^3S_1$ partial wave& $a$  [fm] & $r$  [fm] & $v_2$  [fm$^3$]& $v_3$  [fm$^5$]  & $v_4$  [fm$^7$]   \\
\noalign{\smallskip}\hline\noalign{\smallskip}
NLO KSW  & fit   & fit  & $-0.95$  & $4.6$ & $-25$\\
LO Weinberg  &  fit  & $1.60$  & $-0.05$ & $0.8(1)$  & $-4(1)$\\
Nijmegen PWA & $5.42$  & $1.75$  & $0.04$  & $0.67$ & $-4.0$\\
\noalign{\smallskip}\hline
\end{tabular}
\end{table}
\begin{figure*}
  \includegraphics[width=0.8\textwidth]{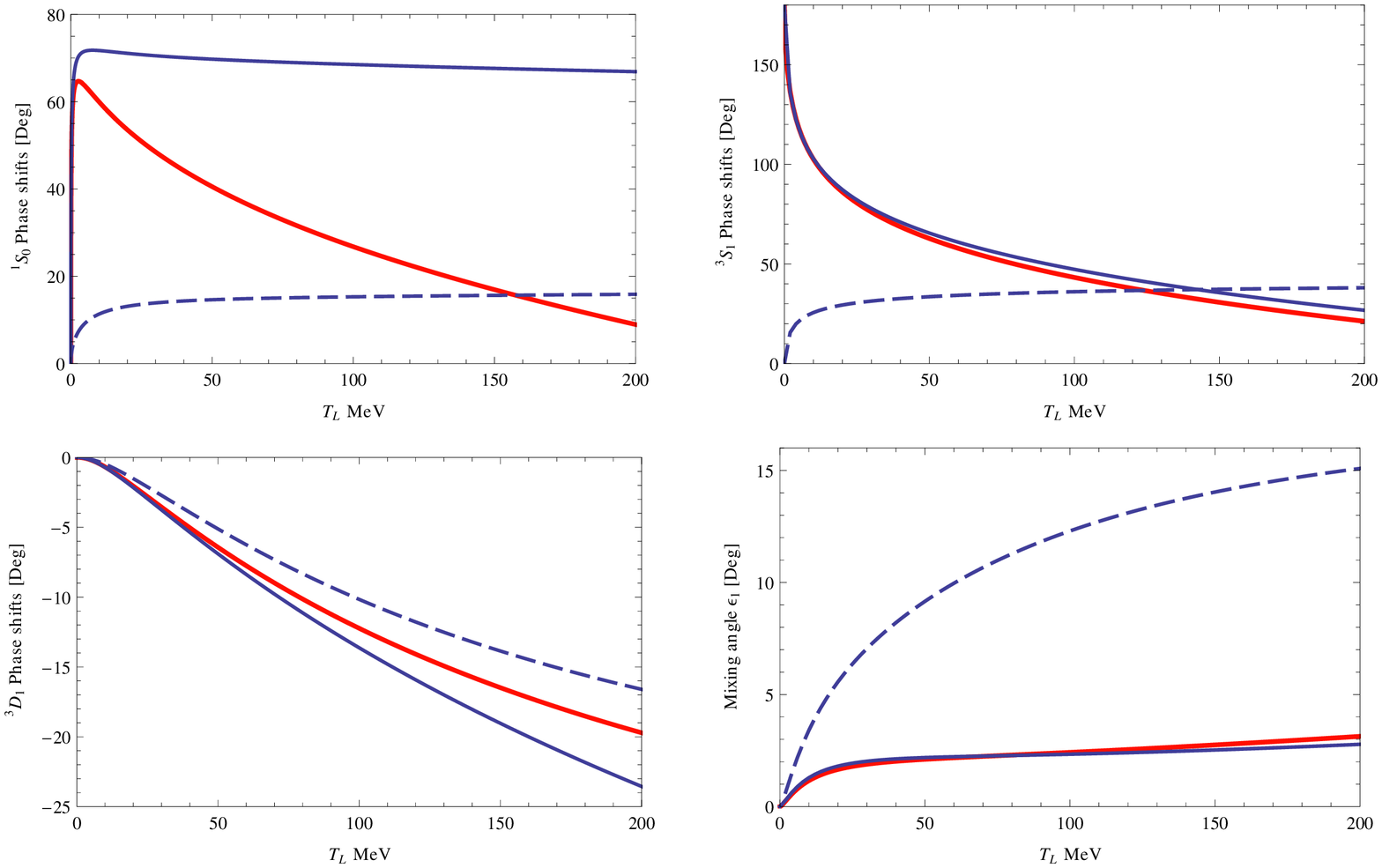}
\end{figure*}
\begin{figure*}
\includegraphics[width=0.8\textwidth]{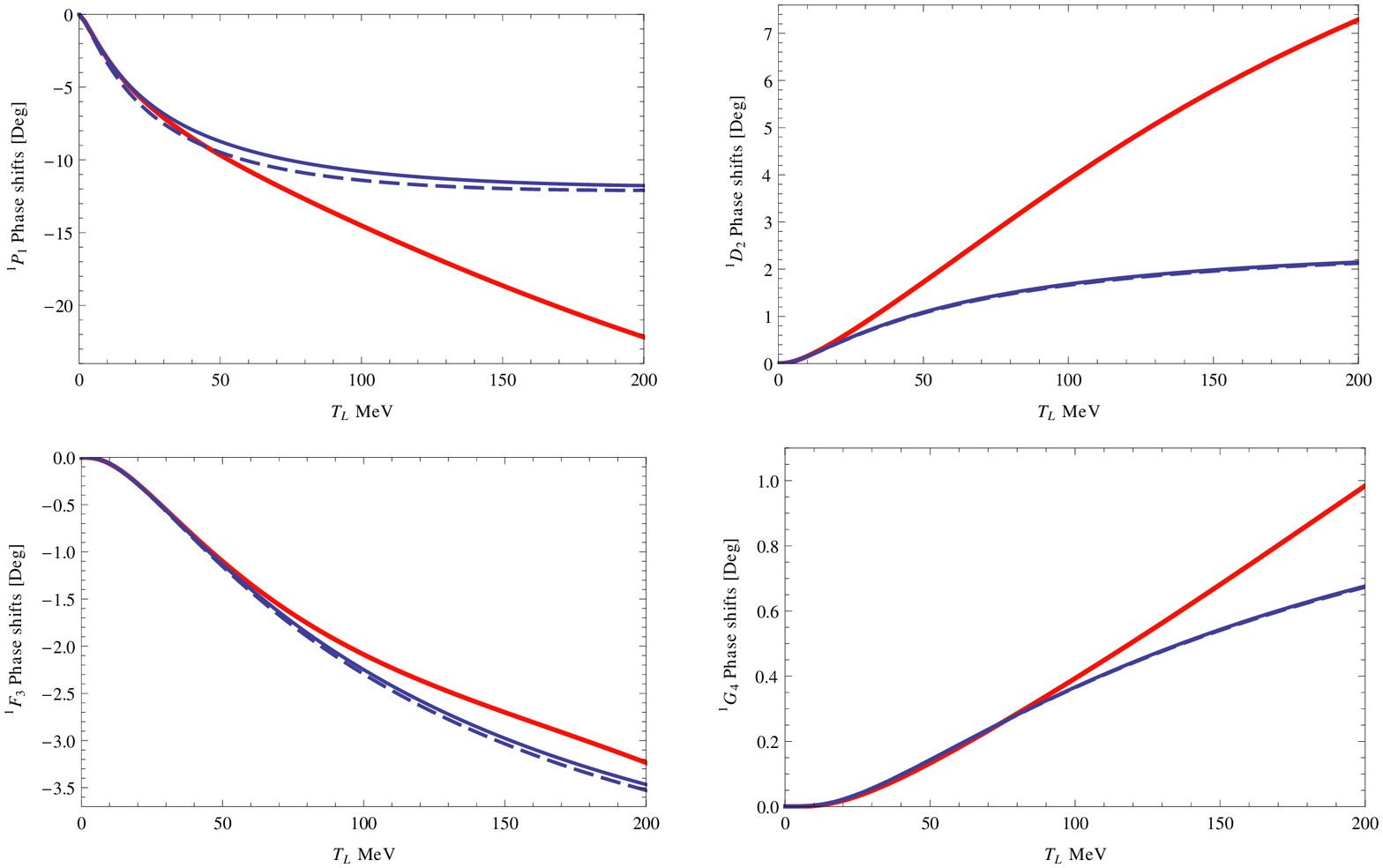}
\caption{Phase shifts at LO. The dashed and solid blue lines and the solid red line correspond to the Born approximation, the solution to the integral equation and the Nijmegen PWA, respectively.}
\label{fig:2}       
\end{figure*}
\begin{figure*}
  \includegraphics[width=0.8\textwidth]{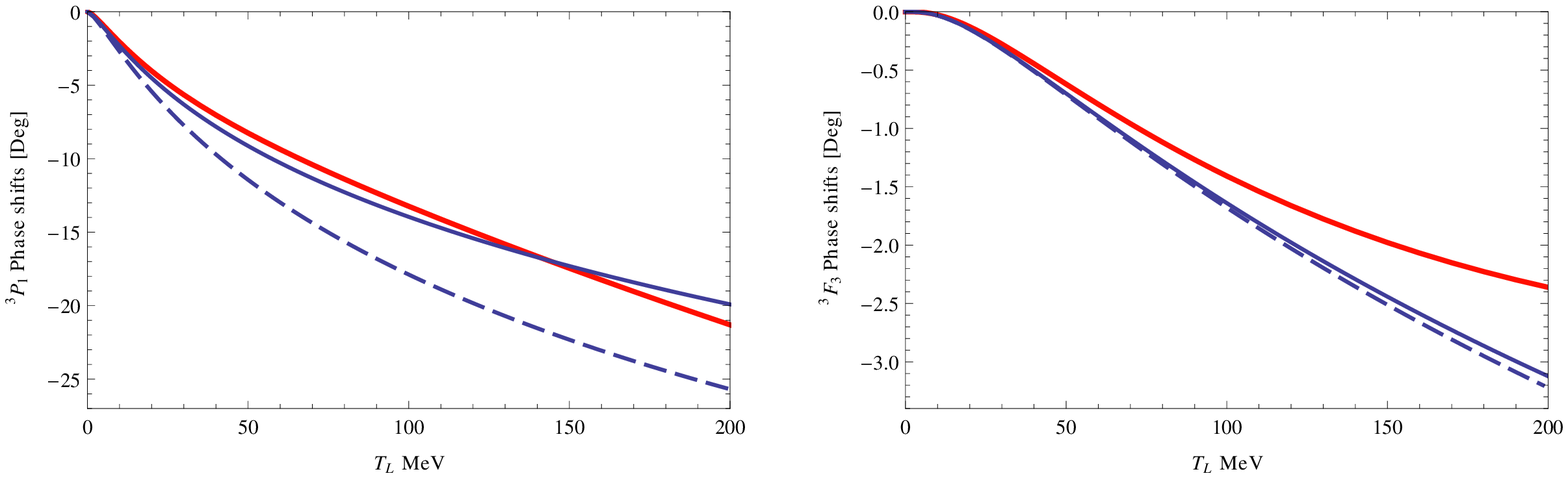}
\end{figure*}
\begin{figure*}
  \includegraphics[width=0.8\textwidth]{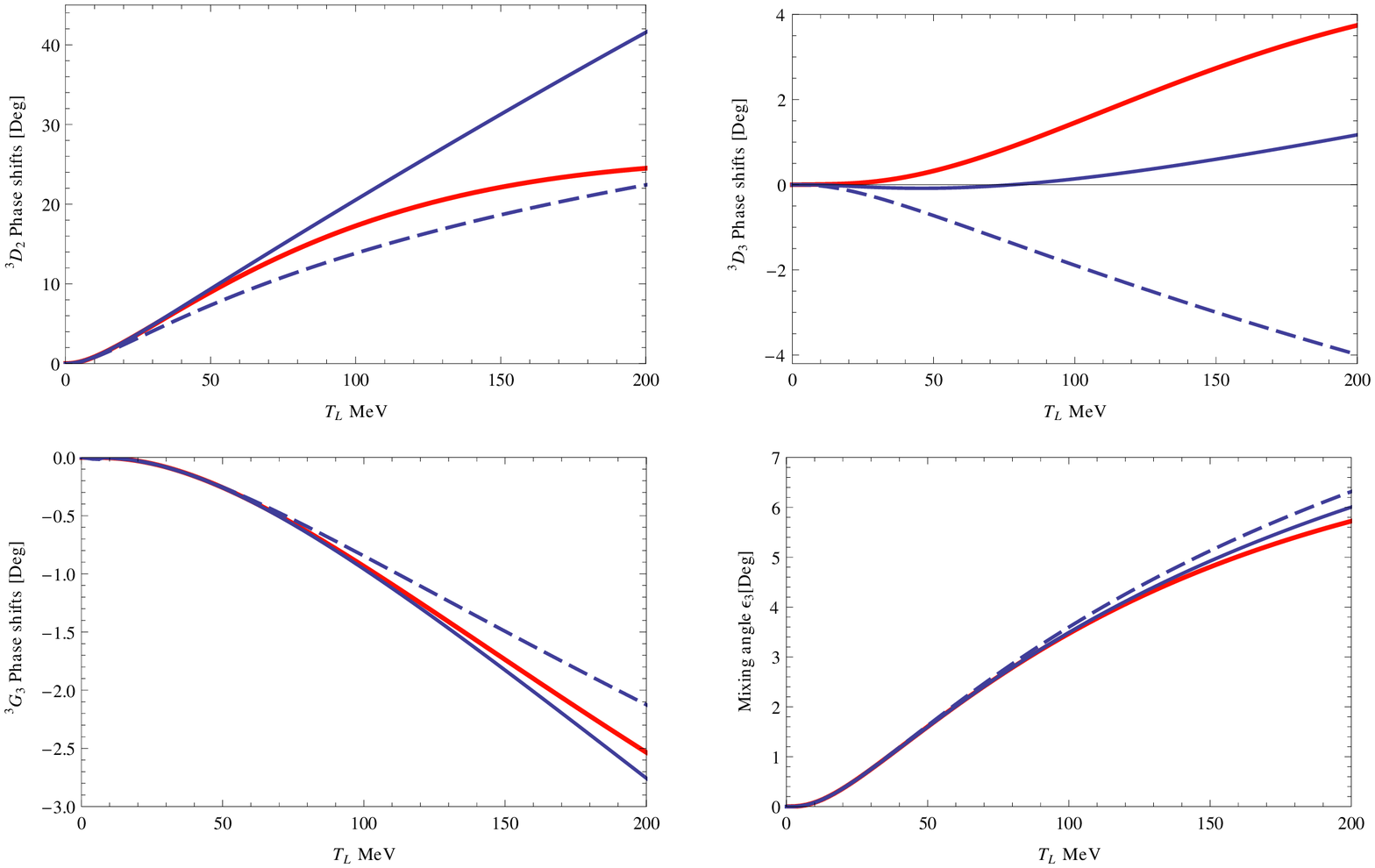}
\caption{Phase shifts at LO. The dashed and solid blue lines and the solid red line correspond to the Born approximation, the solution to the integral equation and the Nijmegen PWA, respectively.}
\label{fig:2}       
\end{figure*}
This work is 
supported by the EU
(HadronPhysics3 project  ``Study of strongly interacting matter''),
the European Research Council (ERC-2010-StG 259218 NuclearEFT)
the DFG (GE 2218/2-1) and
the Georgian Shota Rustaveli National Science Foundation (grant 11/31).

\end{document}